\documentclass{article}
\usepackage{arxiv}

\usepackage[utf8]{inputenc} 
\usepackage[T1]{fontenc}    
\usepackage{hyperref}       
\usepackage{url}            
\usepackage{booktabs}       
\usepackage{amsfonts}       
\usepackage{nicefrac}       
\usepackage{microtype}      
\usepackage{lipsum}		
\usepackage{graphicx}
\usepackage[numbers]{natbib}
\usepackage{doi}
\usepackage{amssymb}
\usepackage{amsmath}
\usepackage{tabu}
\usepackage{tabularx}

\title{Unihedge - a decentralized market prediction platform}

\author{ 
	{\hspace{1mm}Marko Corn,}
	{\hspace{1mm}Nejc Rožman}
}

\renewcommand{\shorttitle}{\textit{arXiv} Template}

\hypersetup{
	pdftitle={Unihedge - a decentralized market prediction platform},
	pdfsubject={q-fin.PR},
	pdfauthor={Marko Corn, Nejc Rožman},
	pdfkeywords={unihedge, hedging, Harberger tax},
}

\begin{document}
	
	\maketitle
	
	\begin{abstract}
		Unihedge is a decentralized platform for prediction markets with a novel approach. Using Harberger Tax (HTAX) economic policies a new type of prediction market, named HTAX prediction market, was build. HTAX prediction market derivates from Dynamic PariMutuel (DPM) type of prediction markets thus offering its users an unlimited liquidity for any preferred time horizon. It tries to solve some problems of DPM by introducing a new incentive mechanism to support early information incorporation and a protection against share readjustment for hedgers. In the paper also implementation of platform on Ethereum Virtual Machine (EVM) is presented with the usage of Decentralized Exchange (DEX) as an price discovery mechanism for prediction market resolutions.
		
	\end{abstract}

	\section{Introduction}
	
	Traditional prediction markets have been centralized in a way where a trustworthy entity maintains a ledger to aggregate trades, similarly the outcome determination of an event and payouts distribution to traders is trusted to a centralized authority \cite{Wang}. However, centralized prediction markets have many risks and limitations: they do not allow global participation, they limit what types of markets can be created or traded, and they require traders to trust the market operator not to steal funds and to resolve markets correctly \cite{Augur}. Similar problems in traditional payment systems have led to the emergence of blockchain technology and cryptocurrencies (e.g. Bitcoin \cite{Bitcoin}). With a decentralized approach, blockchain technology eliminates the centralized intermediary of payments processing and hands over this task to the entire network. Further development of blockchain technology introduced an ability to append logic to transactions which resulted in creation of new type of programmable blockchains (Ethereum \cite{Ethereum}) that were able to run computer programs or so called smart contracts in completely autonomous and trusted way.  
	
	Smart contracts have enabled the decentralization of traditional financial industry by offering decentralized applications for exchanges (Uniswap \cite{Uniswap}), lending (Aave \cite{Aave}), derivatives (Syntetix \cite{Synthetix}), prediction markets (Augur \cite{Augur}, Gnosis \cite{Gnosis}, Polymarket \cite{Polymarket}) and others all united under one ecosystem called DeFi (Decentralized Finance \cite{Chen2019DecentralizedFB}). Example of successful usage of blockchain properties is project named Augur which is a decentralized platform for prediction markets. Augur contracts are totally automated and they hold and transfer users funds, resolve markets and perform other tasks defined by the smart contracts without involvement of developers or any other third parties. 
	
	In this paper we are focusing on prediction markets as a specific part of broader DeFi ecosystem where we present the Unihdge platform with a novel approach to prediction markets. In next section the analysis of existing prediction markets designs are presented, next a concept of HTAX prediction market is presented, flowed by an introduction of decentralized platform Unihedge which implements HTAX prediction market concept for the execution on the Ethereum Virtual Machine (EVM).
	
	\subsection{Prediction markets}
	
	Through the years a variety of financial and wagering mechanisms have been developed for all types of wagering (e.g. sports, horse racing, political events, world news) to support hedging (e.g. insuring) against exposure to uncertain events and speculative trading on uncertain events \cite{Pennock}.  One of those mechanism are prediction markets, which are defined as markets that are run for the primary purpose of information aggregation where market prices forecast future events \cite{Berg}. Prediction markets operate in a form of a market that offers its participants an opportunity to trade outcomes based on their expectations regarding the likelihood of future events. 
	
	Prediction markets operate using contracts, which transform a forecasting outcome into an agreement between participating parties. Contracts are being constructed via trading mechanisms that matches buyers and sellers. There are several trading mechanisms (Table \ref{tab:tradingMechanisms}) among which Continues Double Auction (CDA), Call Auctions (CA), Market Scoring Rules (MSR), parimutuel (PM) and dynamic parimutuel (DPM) are the most commonly used in the field of prediction markets \cite{Luckner}.
	
	\begin{itemize}
		\item CDA is the most commonly used trading mechanism in prediction markets. Traders submit buy and sell orders that are placed on the order book and if matched the contract is made \cite{Madhavan}. One of the main advantages of CDA is that this type of markets pose no risk for the market operator as contracts are made between buyers and sellers \cite{Spann}. However, with few traders the markets may suffer from illiquidity. Orders can then not be matched and therefore the bid-ask spread can be huge and no contract can be created \cite{Hanson}.
		
		\item CA participants place orders to buy or sell units at certain buying or selling prices for a predetermined point in time \cite{Madhavan}. They use single price auctions that match the orders of buyers and sellers, and then, a single trading price is chosen that will maximize volume. CA are more liquid than CDA but limits flexibility for participants.
		
		\item MSR acts like a two-sided market maker that provides infinite liquidity for the sell and buy side of the market with a variable but bounded maximum loss that can be regarded as a subsidy for the market \cite{Hanson}. If traders improve the prediction by moving the prices into the right direction they can expect a positive payoff, otherwise they will lose money. New information is hence reflected immediately \cite{Luckner}. However these market makers are not typically used in real money prediction markets, since taking the opposite side of all trades could result in losses for the market operator.
		
		\item PM markets wagers goes into a central pool and is later divided among the winners \cite{Pennock}. This provides infinite liquidity for wager placement and circumvents the thin market problem of CDA. There is no need for a matching order from another trader. But one shortcoming of parimutuel markets is that there is no incentive to buy contracts early, especially not if new information is expected before the market closes.
		
		\item DPM trading mechanism is a hybrid between the CDA and PM trading mechanism. The DPM offers infinite buy-in liquidity and thus acts as a one-sided market maker always offering to sell at some price and moving the price according to demand. Prices are computed using a price function which can differ depending on the properties that are desired. The DPM also does not exhibit any risk of losses for the market operator due to its redistribution of money. Selling still has to occur through a CDA mechanism because there is no market maker accepting sell offers.
	\end{itemize}
	
	\begin{table}[h]
		\centering	
		\caption{Trading mechanisms}
		\label{tab:tradingMechanisms}
		\begin{tabularx}{\textwidth}{|m{4cm}|m{3cm}|m{1cm}|m{1cm}|m{2cm}|m{2.9cm}|}
			\hline
			& \textbf{CDA} & \textbf{CA} & \textbf{MSR} & \textbf{PM} & \textbf{DPM} \\ \hline
			Continuous information incorporation & \checkmark & $\times$ & \checkmark & $\times$ & \checkmark \\ \hline
			Liquidity & low & higher & infinite & infinite buy in & infinite buy in\\ \hline
			Risk for operator & $\times$ & $\times$ & \checkmark & $\times$ & $\times$ \\ \hline
			Existing decentralized projects & 
			\href{https://augur.net/}{Augur}, 
			\href{https://polymarket.com}{Polymarket}, 
			\href{https://omen.eth.link}{Omen} 
			& / & \href{https://https://gnosis.io}{Gnosis} & / & / \\ \hline
		\end{tabularx}
	\end{table}	
	
	DPM trading mechanism provides infinite buy in liquidity but it is difficult to tell if there is enough incentive for participants to place wagers earlier. Early participants can be deterred from participation in the market in at least two situations: 
	
	First if they are using prediction market as a protection against price volatility of an underling asset. This type of participants, called hedgers, are wagering small amounts on rare unfavourable outcomes. For hedgers the participation in prediction market acts as an insurance, which means they are willing to pay relatively small premium for protection and get sufficiently compensated if unfavourable and rare event occurs. DPM trading mechanism incentivise early placements of wagers with a price function that readjust participants shares in the favour of early participants. Readjustment of shares can pose a risk for hedgers to get devalued to such a degree that their initial wager doesn't provide sufficient compensation. 
	
	The second situation in which early participants can be deterred from participation occurs if they obtain some relevant information about predicted event that will greatly change odds of outcomes. If a participant with this information places wager immediately it reveals it's position which attracts other participants and the readjustment of shares begins. The incentive for the early information incorporation is thus dependent on the price function of DPM which can present a complex calculation.
	
	DPM trading mechanism presents a complex solution for both market operator and participants. Market operator has a challenging task to properly set price function so the market will attract participants and participants have a challenging task to include share readjustment process in their payoff calculations. In this paper we present HTAX prediction market that with the usage of Harberger tax economic policies offers an alternative solution to DMP problems.

	\section{Concept}
	
	Harberger Tax (HTAX) prediction market concept implements HTAX policies in order to provide an environment for the users to trade on the future values of the underlying asset.

	\subsection{Harberger tax policy}
	
	In general HTAX is an economic policy that aims to strike a balance between pure private ownership and total commons ownership of assets in order to increase general welfare of society \cite{Posner}. In this taxation system, asset owners self-assess the value of assets they own and pay a tax rate of on that value. Whatever value owners specify for the asset, they must immediately sell it to anyone at that price. HTAX works as follows:
	\begin{itemize}
		\item Asset owners determines the value of their asset, and pay tax rate on that price. Tax is collected on yearly basis.
		\item Anyone at any time can acquire an asset at the owner’s price, immediately taking ownership of the asset upon purchase. The new owner then sets a new price.
	\end{itemize}
	
	Asset owners are incentivized to set a low price to minimize the amount of taxes they have to pay. At the same time, owners are incentivized to set a price high enough to discourage others from buying it away too easily. This creates a balancing act for owners to price an asset at the value they are willing to pay to keep it \cite{Riady}.
	
	\subsection{Market design}
	
	Prediction markets have to be designed and implemented carefully in order to ensure that they are suitable for aggregating traders’ information \cite{Weinhardt}. The key design elements comprise of specification of contracts, trading mechanism, and the incentives provided to ensure information revelation \cite{Spann}. Implementation of HTAX policies into design of the prediction market follows same structure from contracts to trading mechanism and incentives. 
	
	\subsubsection{Contracts}
	
	Prediction markets participants form binding agreements called contracts with each other that clearly defines agreement conditions and resolutions. HTAX prediction market introduces a standardised product called lot that is part of the contract and can be traded amongst participants. 
	
	\subsubsection*{Lot} \label{sec:lot}
	
	Lot derives from the concept of virtual land $L$ which is a two dimensional set with one dimension being time $t$ and other being outcome domain of the observed event of the prediction market $o_d$ (example: future price of an asset) as defined in Equation \ref{equ:land}.
	
	\begin{equation}
		\mathcal{L}={(t,o_d)\in \mathbb{R} \times \mathbb{R}}
		\label{equ:land}
	\end{equation}
	
	Land is divided into smaller pieces called lots that can be traded between users. Lots are subjected to the HTAX policies that transforms them into assets. Lots are formed by discretization of virtual land in both dimensions. Time discretization (sampling) with fixed period ($period$) divides time dimension into equal parts called frames. Each frame is defined by start and end time as shown in Equation \ref{equ:frames}.
	
	\begin{equation}
		frame_n=[k_{n},k_{n}+period]
		\label{equ:frames}
	\end{equation}
	
	The discretization (quantization) of the event outcome domain divides it into discrete sets of outcomes with top and bottom limits as shown in Equation \ref{equ:sets}. $\Delta o$ represents granularity of quantization process. 
	
	\begin{equation}
		outcome_m=[o_m, o_m+\Delta o]
		\label{equ:sets}
	\end{equation}
	
	Lot is defined as a container with four barriers, two of time and two of outcome as shown in Equation \ref{equ:block}.
	
	\begin{equation}
		block_{n,m}=[k_n, k_{n}+period, o_m, o_m+\Delta o]
		\label{equ:block}
	\end{equation}
	
\begin{figure}[h]
		\centering
		\includegraphics[scale=0.7]{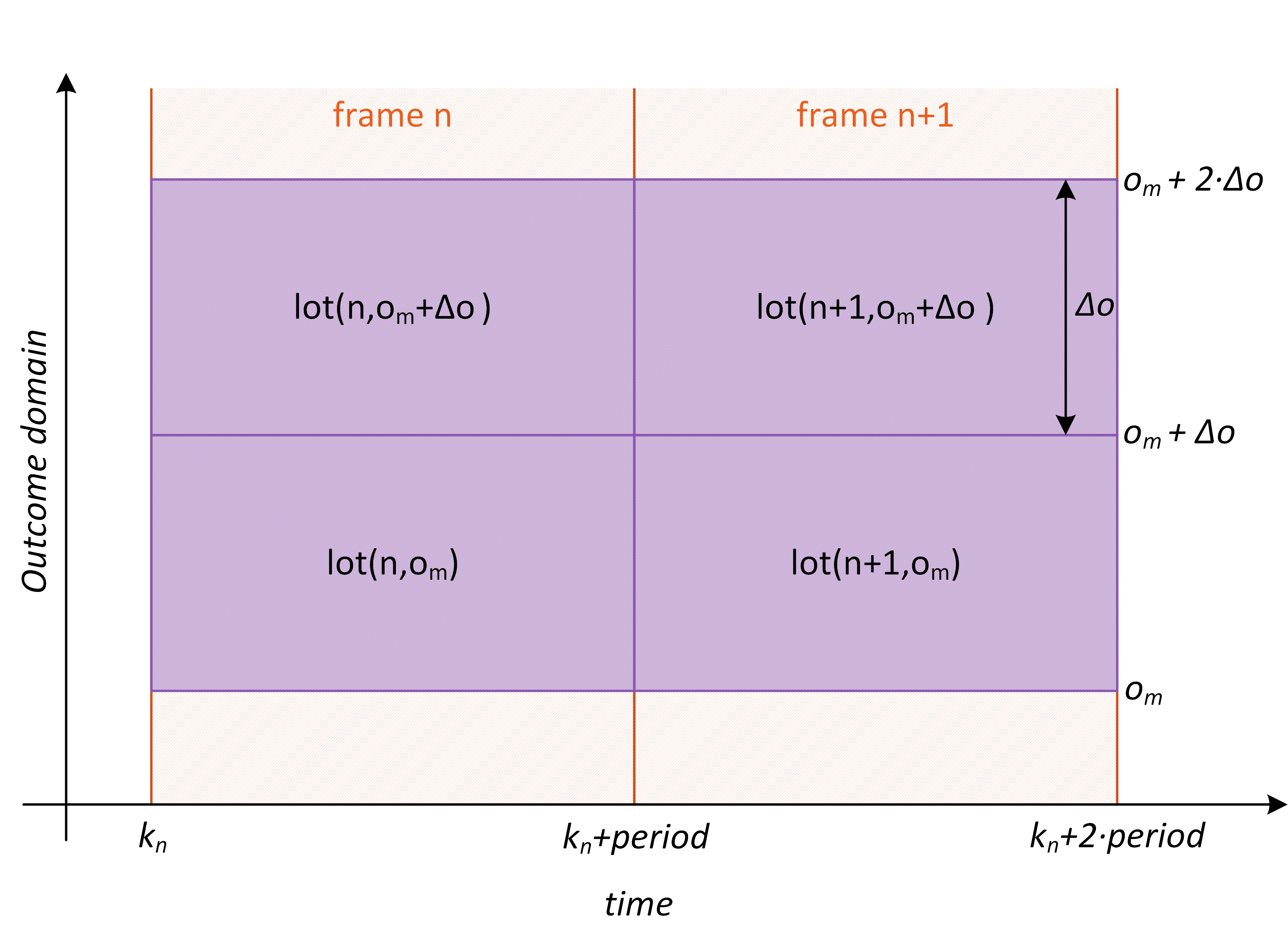}
		\caption{Representation of lots}
		\label{fig:lots}
	\end{figure}

	An example of lots is shown in Figure \ref{fig:lots}.
	
	\subsubsection*{Life cycle}
	
\begin{figure*}[h]
		\centering
		\includegraphics[width=\linewidth]{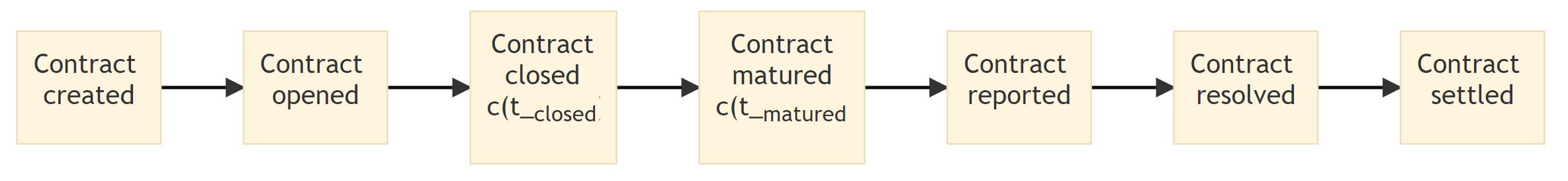}
		\caption{Contract's life cycle.}
		\label{fig:contractLifeCycle}
	\end{figure*}

	A contract has several states that comprise its life cycle as shown on Figure \ref{fig:contractLifeCycle}. 
	
	\begin{itemize}
		\item \textbf{Created}: Contract $c(t_{created})$  is created at time $t_{created}$ and at same time the contract's outcome event is defined.
		\item \textbf{Opened}: Contract in opened state allows trading of the lots.
		\item \textbf{Closed}: Contract in closed state has ended trading at time $t_{closed}$ which means no lot can be resold.
		\item \textbf{Matured}: Contract is matured when predicted event occurs at time $t_{matured}$ where $t_{matured}>t_{closed}$. After maturation of the contract the reporting of the outcome of predicted event begins.
		\item \textbf{Reported}: Outcome is reported to the contract via arbitrary oracle system
		\item \textbf{Resolved}: Resolvement of the contract determines winning lot and calculate shares of the reward for current and past owners of the winning lot.
		\item \textbf{Settled}: Contract in settled state has distributed winnings to the users.
	\end{itemize}
	
	\subsubsection{Trading mechanism}
	
	Trading mechanism enables creation and modification of contracts. In HTAX prediction market each frame can hold one contract which means a contract can hold only lots that belongs to the same frame ($\{lot_{n,0},_{n,1},...,lot_{n,m}\}$). Trading mechanism enables two main contract process: lot trading and tax collection.
	
	\subsubsection*{Lot trading}
	
	Lot trading is a process where lots change ownership. Lot can be purchased (change ownership) if the acquisition price, set by the current owner, is matched by another trader. When the lot is purchased the new owner sets new acquisition price for the lot. If lot has no owner the starting acquisition price of the lot is zero.  
	
	\subsubsection*{Tax collection}
	
	Tax collection is a process where each owner gets charged with a tax proportional to the acquisition price and duration of the ownership for the lot as shown in Equation \ref{equ:tax}.
	
	\begin{equation}
		tax=\frac{rate}{period}\cdot price_{lot}
		\label{equ:tax}
	\end{equation}
	
	Rate ($rate$) and ($period$) form a tax rate (example: 1\%/year) to be collected from the owners of the lots. All tax collected per lot is collected in a pool ($pool_n$) and is equal to the sum of taxes collected from all owners ($O$) until contract is closed as shown in Equation \ref{equ:taxTotal}. $span_i$ represents the duration of the ownership over lot by owner $i$, $price_i$ is set price for lot by owner $i$ and $n$ is the frame's number.
	
	\begin{gather}
		pool_n=\sum_{i=0}^{O}rate\cdot price_i \cdot span_i
		\label{equ:taxTotal}
	\end{gather}
	
	\subsubsection{Incentives}
	
	Incentives represents incomes for the users of the platform. There are two type of incentives a direct incentives in the form of rewards and an indirect incentive in the form of lot trading.  
	
	\subsubsection*{Direct incentives}
	
	Direct incentives are rewards that are provided by the frame's pool. The amount for rewards in a pool is equal to the total tax collected according to Equation \ref{equ:taxTotal}. Each frame has one winning lot to which whole pool is assigned. Lot has won if outcome value sits inside lot's outcome set \ref{equ:outcome}.
	
	\begin{equation}
		lot_{n,m} =
		\begin{cases}
			won; & \quad \text{if } o_m<=outcome\ \cap \  o_{m+\Delta}>outcome\\
			lost;  & \quad \text{if } o_m>outcome\ 	\lor \  o_{m+\Delta}<=outcome
		\end{cases}
		\label{equ:outcome}
	\end{equation}
	
	The amount in a pool assigned to the winning lot belongs to the lot's current owner. 
	
	\subsubsection*{Indirect incentives}
	
	Indirect incentives represents an income opportunity for the users to gain on the future price movements of the lots (speculation). For example users can buy cheaper lots further in time where uncertainty for those lots to win is higher and later sell them when uncertainty is lower.
	
	\subsection*{Concept summary}
	
	HTAX prediction market concept introduces a new type of property called lot which is a part of a virtual land that represents all future values of observed variable. The ownership of lots prevents share readjustment process of DPM as only way to acquire lot is to purchase it from the current owner which sets its selling conditions. This feature benefits the hedgers and participants with new information. 
	
	It benefits hedgers when a hedger purchase a lot with low probability of winning for low price. Their acquisition price should be set high enough that selling it would compensate their loss in the underling asset. In the case where the lot won't be sold and would won, the hedger benefits from the reward.     
	
	It benefits participants with new information when participant with a new information, sets price for the revelling information by setting acquisition price for the lot. This stimulates participants with new information to purchase the lot as soon as the information arrives and by doing so gain an advantage over other participants.
	
	One more advantage of HTAX prediction market is that at the time of lot purchasing users knows minimum payoff amount that this purchase will generate. At the time of the lot purchase the current reward pool amount and tax that needs to be paid are known. From this information the buyer can calculate his minimum reward in case no one purchase his lot or sets the acquisition price that provides him some gain. This property offers additional simplification for participation in this type of prediction market.     
	
	To conclude, HTAX prediction market eliminates the risk for the market creator to correctly set a price function, relieves participants of complex calculation of their payoff, it prevents share readjustment process for the hedgers and incentivises early information incorporation. 
	
	\section{Decentralized platform}
	
	Unihedge is a decentralized platform build as a fully decentralized application that can run on any EVM compatible blockchain. It implements HTAX prediction market concept in the form of smart contracts. In the forthcoming section the definition of the market parameters is presented, followed by description of market creation and operation processes.   
	
	\subsection{HTAX prediction market}
	
	Implementation of HTAX prediction market as smart contract requires parameters that are listed in the table \ref{tab:marketPar}.
	
	\begin{table}[h]
		\centering
		\caption{List of parameters of HTAX prediction market implementation}
		\begin{tabu}{|l|l|l|l|}
			\hline
			\textbf{Name} & \textbf{Type} & \textbf{Description} & \textbf{Unit} \\ 
			\hline\tabucline[1pt]{-}
			accounting token & ERC20 & token for lot trading & / \\ \hline
			trading pair & address & trading pair address from the exchange & / \\ \hline
			granularity & integer & outcome span of lots & / \\ \hline
			initial timestamp & integer & debut time of market & Unix timestamp \\ \hline
			period & integer & duration of frame & seconds \\ \hline
			reporting interval & integer & interval for calculation of average rate & seconds \\ \hline
			tax & integer & tax that is charged to the owners of lots & \% \\ \hline
			market fee & integer & fee that is collected by the market operator & \% \\ \hline
			protocol fee & integer & fee that is collected by the the platform  & \% \\ \hline
		\end{tabu}
		\label{tab:marketPar}
		
	\end{table}
	
	\textit{accounting token} is an ERC20 token that is used for lot trading, \textit{trading pair} address represents a Decentralized Exchange (DEX) market that provides rate discovery for a pair of ERC20 tokens, \textit{granularity} sets the outcome span of the lots based on definition of lot in section \ref{sec:lot}, \textit{initial timestamp} sets debut time for market operations (before this time market is not operational), \textit{period} sets the frames duration based on definition of the frame in section \ref{sec:lot}, \textit{reporting interval} defines averaging window size for the calculation of outcome value as defined in section \ref{sec:reporting}, \textit{tax} defines tax rate that is collected from the owners of the lots, \textit{market fee} and \textit{protocol fee} is fee that is collected from the users. 
	
	\subsection{Market creation} \label{sec:marketCreation} 
	
	The creation of a new HTAX prediction market can be done by anyone. There is no limit to the number of markets that can be created nor any limits regarding market duplicates.

	\subsection{Market operation}
	
	Market operation consists of several sub processes:
	
	\begin{itemize}
		\item \textbf{automatic contract creation} enables creation of contracts that are basis for trading lots,
		\item \textbf{trading lots} which enables users to exchange lots ownership,
		\item \textbf{tax collection} which is responsible to collect tax from lot owners,
		\item \textbf{reporting on rate} which enables extraction of rate for token pair from the DEX,
		\item \textbf{resolvement} which defines winners and losers and calculate rewards,
		\item \textbf{settlement} which enables winners to collect the rewards.
	\end{itemize}	
	
	\subsubsection{Automatic contract creation}
	
\begin{figure*}[h]
		\centering
		\includegraphics[scale=0.75]{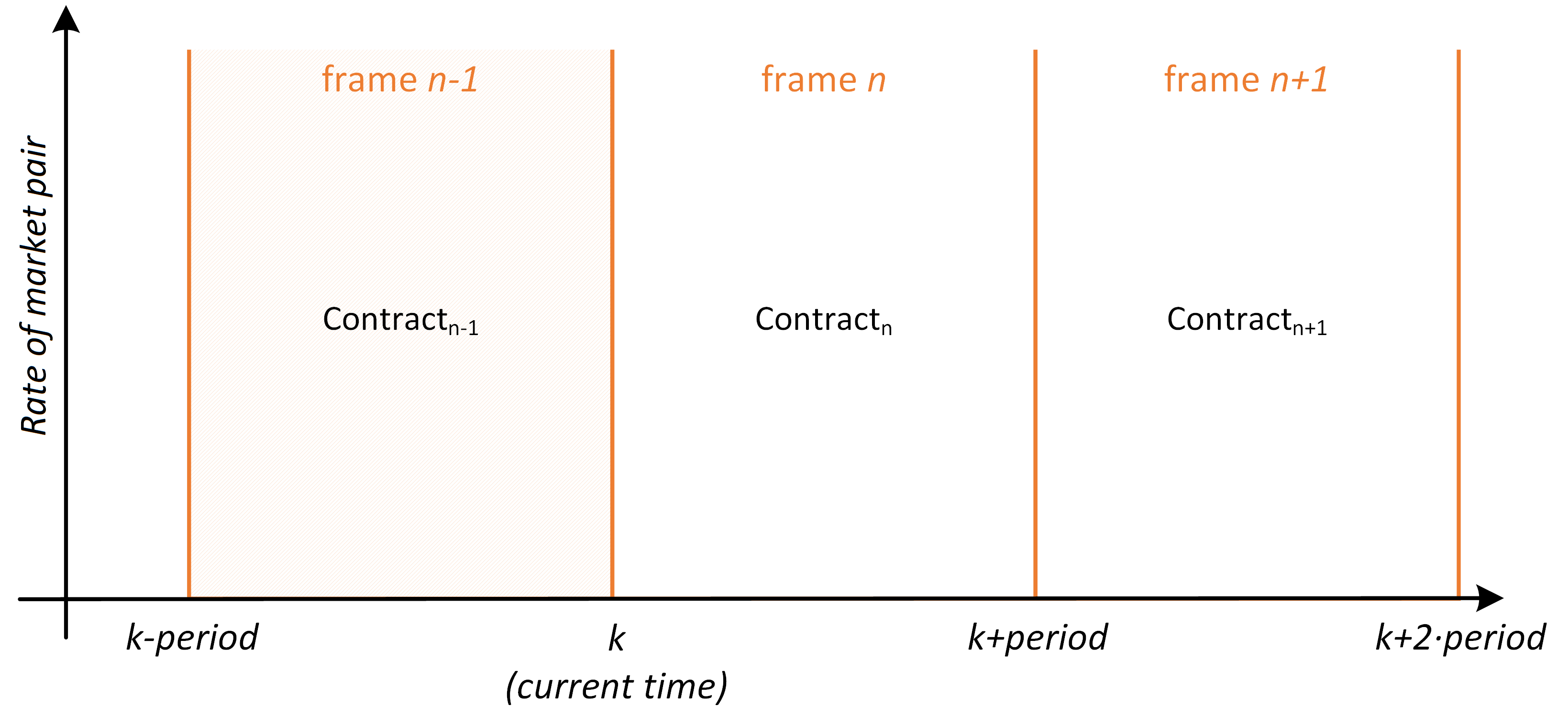}
		\caption{Automatic contract creation}
		\label{fig:acc}
	\end{figure*}

	The process of automatic contract creation begins with a discretization of the asset's rate timeline with a period ($period$) (e.g. 24 hours, 7 days...) as is it shown in Figure \ref{fig:acc}.
	
	Discretization breaks the asset's rate timeline into frames $f_i$ each belonging to one contract $c_i$ (Equation \ref{equ:frameDef}).
	
	\begin{gather}
		f_i\in c_i\nonumber\\
		f_i=[k_{start}, k_{end}=k_{start}+period]
		\label{equ:frameDef}
	\end{gather}
	
	The state of the contract is depended on the current time. Looking at Figure \ref{fig:acc}, the contract $c_{n-1}$ is matured in current time, which means that predicted event has occurred and asset's rate can be extracted from the DEX. Contract $c_{n-1}$ can now be in reported, resolved or settled state. Contract $c_{n}$ is in closed state because it is not yet matured and it does allow lot trading. Contract $c_{n+1}$ and any other subsequent contracts $c_{n+1+m}$ are in open state allowing lot trading.  
	
	Process of automatic contract creation creates new contract when user wants to purchase lot within a valid frame where no lot has been previously bought. A valid frame represents any frame that starts at time greater than the current time (Equation \ref{equ:frameValid}). 
	
	\begin{equation}
		k_{start}>k
		\label{equ:frameValid}
	\end{equation}
	
	\subsubsection{Trading lots and tax collection}
	
\begin{figure*}[h]
		\centering
		\includegraphics[scale=0.45]{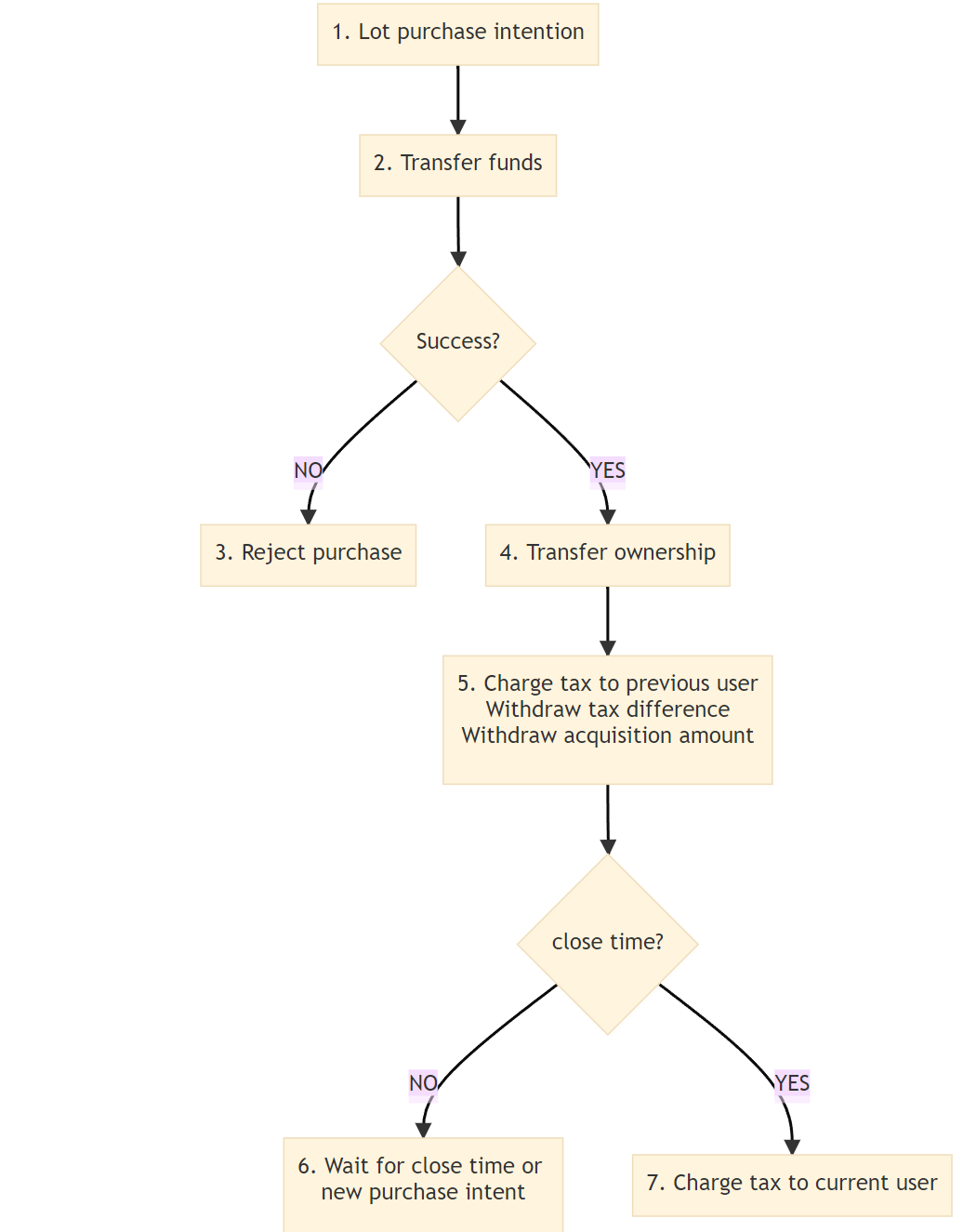}
		\caption{Process of lot trading and tax collection}
		\label{fig:processLotTrading}
	\end{figure*}

	The prediction market module enables the exchange of lot ownership between users. When a user decides to purchase a lot, it selects a lot within a valid frame. If the lot has no owner, the acquisition price for it is zero, otherwise the user has to pay acquisition price set by the current owner. Tax collection process is responsible to collect tax from owners of lots. It is presented together with trading process as they are interconnected as show on diagram on Figure  \ref{fig:processLotTrading}. Lot purchase starts with step 1. In this step the maximum tax is calculated as if the owner will own lot until close time of lot's frame as shown in Equation \ref{equ:maxtax}.
	
	\begin{equation}
		tax_{max}=rate\cdot (t_{close}-t)
		\label{equ:maxtax}
	\end{equation}	
	
	$tax_{max}$ is equal to the product of $rate$ and duration between current time $t$ and close time $t_{close}$. In step 2 funds are transferred in custody to contract. Funds represents cost of purchasing lot that consists of maximum tax ($tax_{max}$) and acquisition amount ($amount_{aqu}$) for the current owner of the lot if there is one as shown in Equation \ref{equ:funds} 
	
	\begin{equation}
		cost=tax_{max}+amount_{aqu}
		\label{equ:funds}
	\end{equation}
	
	Step 4 transfer ownership of lot to new owner if transfer of funds was successful and in other case lot purchase process is rejected. In step 5 tax for owning lot is calculated and charged to the previous owner as shown in Equation \ref{equ:taxCharged}.
	
	\begin{equation}
		tax_{charged} = rate \cdot span = rate \cdot (t-t_{acquisition})
		\label{equ:taxCharged}
	\end{equation} 
	
	$tax_{charged}$ is equal to the product of $rate$ and $span$ of ownership that is defined as difference between current time $t$ and time of lot acquisition $t_{acquisition}$. Tax is deducted from the contract custody fund and transferred to frame's $pool$ (Equation \ref{equ:taxTotal}) where it will be distributed as rewards to owners of the winning lot. The previous owner is now allowed to withdraw the acquisition amount paid by the new owner of the lot and difference between $tax_{max}$ and $tax_{charged}$. in step 7, after the close time of the frame all current owners of lots are charged with tax as no new acquisitions can be performed.    
	
	\begin{figure*}[h]
		\centering
		\includegraphics[scale=0.5]{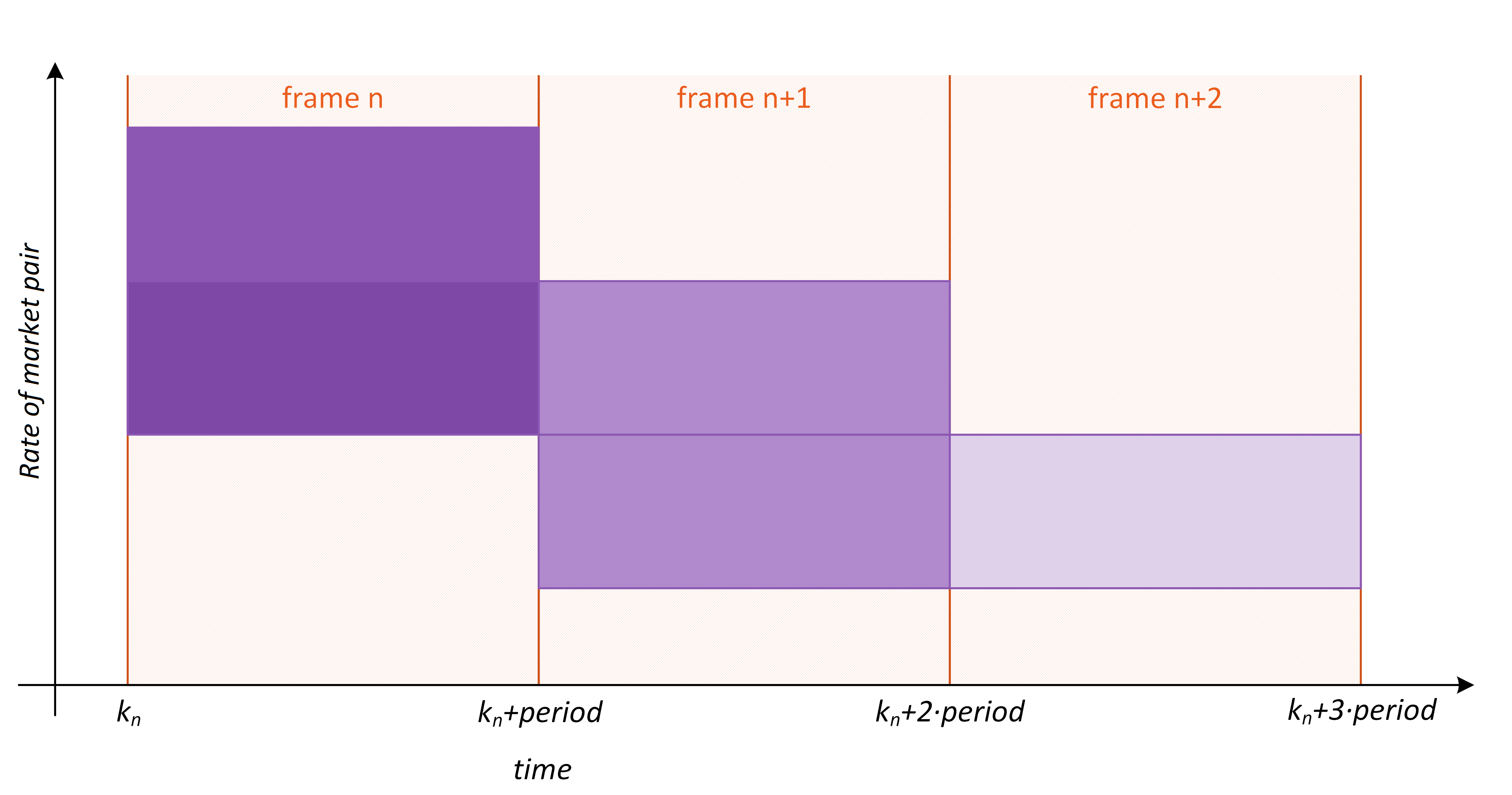}
		\caption{Example of purchased lots}
		\label{fig:placedwagers}
	\end{figure*}

	\subsubsection{Reporting on rate} \label{sec:reporting}
	
\begin{figure*}[h]
		\centering
		\includegraphics[scale=1]{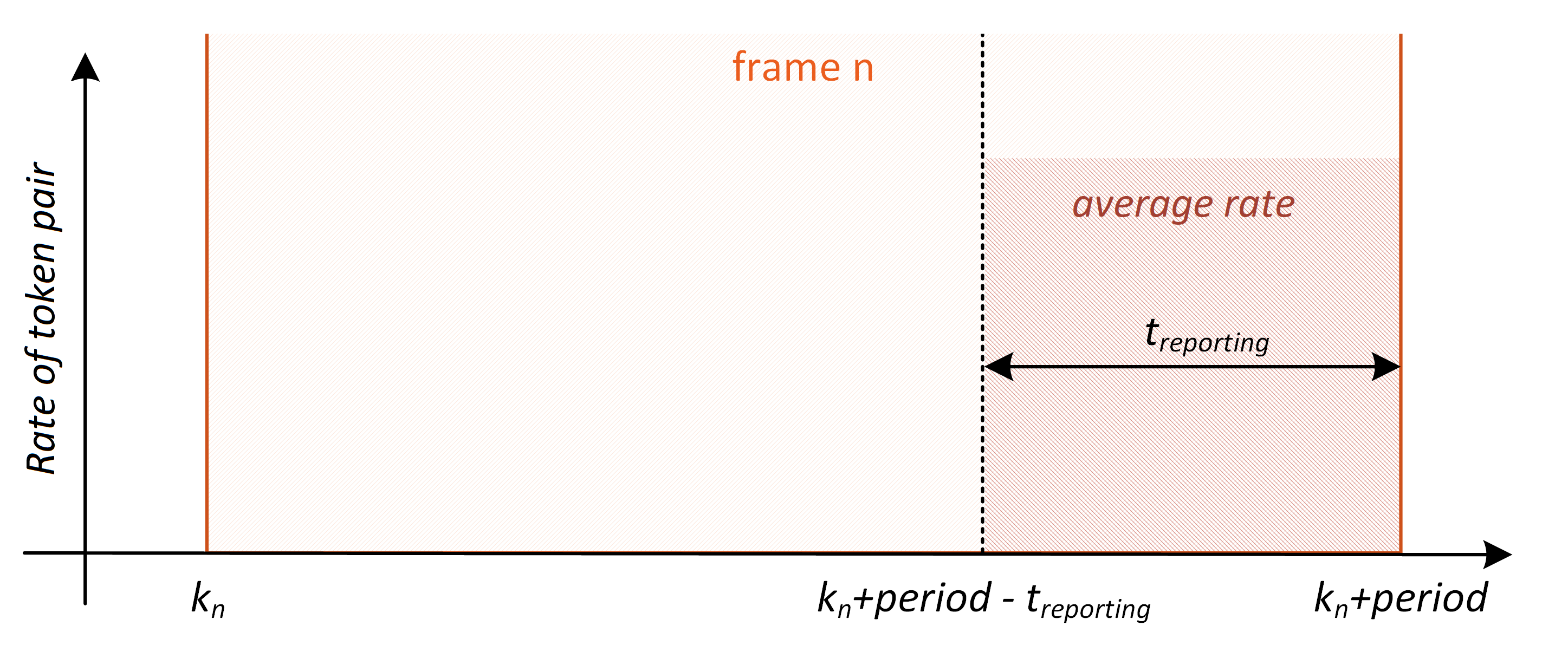}
		\caption{Reported rate}
		\label{fig:reportingRate}
	\end{figure*}

	The rate $rate_{reported}$ that is reported for the resolution of the contract is equal to the average rate of token pair that spans from close time $k_{n}+period$ of the frame $n$ to the close time of the frame reduced by reporting interval $t_{rep}$ as show on Figure \ref{fig:reportingRate}. The reporting interval is set by the market creator at the creation of the market.
	
	The platform relies on the usage of DEXs as rate oracle to provide accurate rate feeds for the contract resolutions. It utilizes a special type of DEX that uses Automated Market Maker (AMM) as rate discovery mechanism with integrated oracle for obtaining average rates. Examples of this type of DEX are Uniswap \cite{Uniswap} deployed on Ethereum network or Honeyswap \cite{Honeyswap} deployed on XDAI network. Reporting on rate is done via this integrated oracle system that enables calculation of time weighted average rates (TWAP) of a token pair. TWAP is constructed by obtaining two snapshots of cumulative rates, first cumulative rate $rate_{cum\ 1}$ of a token pair at the beginning $t_1$ and second cumulative rate $rate_{cum\ 2}$ at the end $t_2$ of the observed interval. The difference in cumulative rates can then be divided by the length of the interval ($t_2-t_1$) to create a TWAP for that time span as shown in Equation \ref{equ:rateUniswap}. 
	
	\begin{equation}
		TWAP=\frac{rate_{cum\ 2} - rate_{cum\ 1}}{t_2-t_1}
		\label{equ:rateUniswap}
	\end{equation}
	
	Using TWAP to calculate $rate_{rep}$ requires to change $t_1$ timing with $k_{n}+period-t_{reporting}$ and $t_2$ with $k_n+period$ as shown in Equation \ref{equ:rateUniswap2}. 
	
	\begin{equation}
		rate_{rep}=\frac{rate_{cum\ 2} - rate_{cum\ 1}}{k_n+period-(k_{n}+period-t_{rep})}=\frac{rate_{cum\ 2} - rate_{cum\ 1}}{t_{rep}}
		\label{equ:rateUniswap2}
	\end{equation}
	
\begin{figure*}[h]
		\centering
		\includegraphics[scale=1]{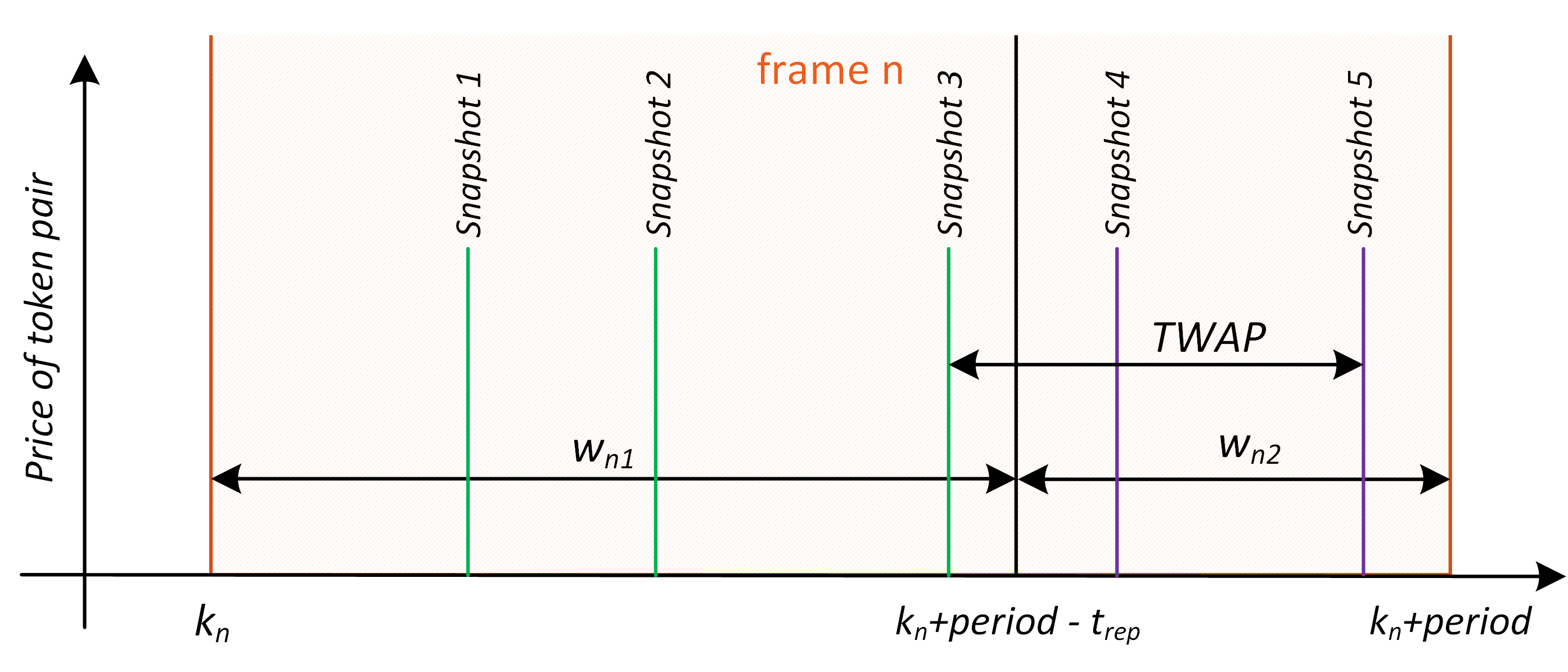}
		\caption{Timing of snapshots}
		\label{fig:reportingTwap}
	\end{figure*}

	The rate reporting can't be executed internally by smart contract as smart contract can't execute transaction by themselves it must be executed externally by users \cite{Ethereum}. Since rate reporting is done by users of the platform the exact timing for rate reporting can't be guaranteed. A solution to this problem is a process that mitigate the exact timing and offers time windows for reporting on rate. The reporting process offers two windows one for each snapshot of oracle's cumulative rate. First window $w_{n_1}$ spans from $k_n$ to $k_n+period-t_{rep}$ and second window $w_{n_2}$ spans from $k_n+period-t_{rep}$ to $k_n+period$. If user triggers reporting event at time that corresponds to any of the windows the snapshot values are stored by the platform for calculation of $rate_{rep}$. In case of multiple reporting events the platform override old snapshot with new one. This process guarantees that average rate will take into account $t_{rep}$ interval as shown on Figure \ref{fig:reportingTwap}. Example on Figure \ref{fig:reportingTwap} shows that TWAT was calculated from snapshot 3 and 5 as they were the last snapshots taken for each window.
	
	Users can trigger rate reporting event in two ways:
	
	\begin{itemize}
		\item indirectly, with the purchase of a lot that belongs to some other frame in the future
		\item directly, by invoking platforms function. This is an additional option that allows the user to trigger reporting event without trading process. Direct way of triggering reporting process is facilitated by the users which expect to win. 
	\end{itemize}

	\subsubsection{Resolvement}\label{sec:resolvement}
	
	In the resolvement phase the owner of the winning lot is set and the reward is calculated as the amount collected in a pool ($pool_n$) minus fee as shown in Equation \ref{equ:reward}.
	
	\begin{equation}
		\label{equ:reward}
		reward_i=share_{lot_i} \cdot (pool-fee)
	\end{equation}
	
	If an unsuspected event occur and this calculation is not possible, the contract goes into invalid state which means the funds will be returned to the users except the collected fees. Invalid state can occur if:
	
	\begin{itemize}
		\item no lot was purchased from specific contract
		\item there is no winning lot
		\item an error has occurred in reporting process
	\end{itemize}
	
	\subsubsection{Settlement}
	
	Once the resolvement process is completed, the winner can withdraw the reward by executing a settlement process. The reward funds of the user remains locked on the contract until a withdrawal is performed by the user. Also at this stage, users can withdraw funds from contract that are in invalid state.

%
%
%
%
%
%
%
%
%
%
%
%
%

	\section{Conclusion}
	
	We presented a decentralized platform named Unighedge with it's novel approach to prediction markets. Using HTAX economic policies a new type of prediction marketg, named HTAX prediction market, was built. HTAX prediction market offers: 
	
	\begin{itemize}
		\item an unlimited trading with infinite buy-in liquidity and for any preferred time horizon,
		\item a novel incentive mechanism to support early information incorporation,
		\item a protection against share readjustment for hedgers,
		\item relieves market creators of complex setting of price function,
		\item relieves participants of complex calculation of their payoff.  
	\end{itemize}
	
	\bibliographystyle{unsrt}
	\bibliography{whitepaper}

\end{document}